\documentclass[aps,twocolumn,groupedaddress,floatfix,amsmath,amssymb,pra,longbibliography]{revtex4-1}
\usepackage{amsmath}
\usepackage{amssymb}
\usepackage{graphicx}
\usepackage{textcomp}
\usepackage{bm} 
\usepackage{braket} 
\usepackage{longtable}
\usepackage[usenames,dvipsnames]{color}

\begin{document}

\title{Slow-light frequency combs and dissipative Kerr solitons in coupled-cavity waveguides}

\author{J. P. Vasco}
\affiliation{Institute of Theoretical Physics, Ecole Polytechnique F\'ed\'erale de Lausanne EPFL, CH-1015 Lausanne, Switzerland}
\author{V. Savona}
\affiliation{Institute of Theoretical Physics, Ecole Polytechnique F\'ed\'erale de Lausanne EPFL, CH-1015 Lausanne, Switzerland}

    
\begin{abstract} 
We study Kerr frequency combs and dissipative Kerr solitons in silicon photonic crystal coupled-cavity waveguides (CCW) with globally optimized dispersion at telecom wavelengths. The corresponding threshold for comb generation is found to explicitly depend on the main CCW figures of merit, namely, mode volume, normal mode quality factor and slow-light group index. Our analysis is carried out by solving the non-linear dynamics of the CCW Bloch modes in presence of Kerr non-linearity and two-photon absorption. Our results open the way to CCW comb generation via dispersion engineering and slow-light enhancement.
\end{abstract} 

\maketitle

\section{Introduction}

Kerr frequency combs have revolutionized several fields in optical sciences since they were first proposed in monolithic microresonators \cite{kippenberg0}. To date, they have been successfully applied to a vast variety of state-of-the-art technologies and highly sophisticated measurement techniques as ultra-high resolution spectroscopy \cite{swann,hansch}, massively parallel coherent optical communications \cite{koos1}, light detection and ranging (LIDAR) \cite{koos2}, optical-frequency synthesizers \cite{scott} and microphotonic astrocombs \cite{benedick,herr}. Frequency combs in optical microresonators are generated by the interaction of {either, a continuous-wave (cw) pump laser or pumped optical pulses \cite{herr2},} with the resonator modes via parametric four-wave mixing (FWM), assisted by the Kerr non-linearity of the material. This parametric process fulfills the energy conservation and is enhanced when the side bands created by FWM coincide with the resonator frequencies. Since the intensity-dependent refractive index induces a frequency shift on the modes, an increasing free spectral range (FSR) with frequency (anomalous dispersion) is required to compensate this effect and effectively produce equidistant spectral lines suitable to support the cascaded FWM \cite{kippenberg5}. When a high number of mode-locked modes are involved in this non-linear interaction and dissipation cannot be neglected, the complex non-linear dynamics may give rise to disipative Kerr solitons (DKS), which arise as a double balance between non-linearity and dispersion (preserving their shape), as well as dissipation and parametric gain (preserving their amplitude) \cite{kippenberg4}. {Moreover, multiple-soliton formations have become very interesting in the context of localized-state interactions \cite{tlidi,rivas2,torner}, breather solitons \cite{gaeta1,kippenberg6,kippenberg7} and soliton crystals \cite{cole,karpov}.} DKS are notably relevant because their corresponding spectra exhibit highly coherent frequency combs with perfect single-FSR spacing between side bands \cite{kippenberg1,gaeta5}, and dispersive waves in presence of high-order dispersion \cite{erkintalo,skryabin,rivas,kippenberg2}. The latter is particularly important because it allows a further spanning of the frequency combs, highlighting the important role of dispersion engineering on parametric gain, and consequently, on comb generation \cite{gaeta2}. Such task has been successfully achieved in non-linear microresonators by varying the ring geometrical parameters in order to tailor the dispersion for specific comb functionalities \cite{michel,gaeta3,yu}. Nevertheless, this geometry has a limited parameter space, thus restricting the choice of materials, operation wavelength and size of the final device. Recently, Fabry-Perot resonators have been proposed as appealing candidates for frequency comb generation since different methods to reshape the cavity dispersion can be applied \cite{lugiato,papp}, however, advanced dispersion engineering still remains challenging. 

\begin{figure}[t!]
\centering
\includegraphics[width=0.47\textwidth]{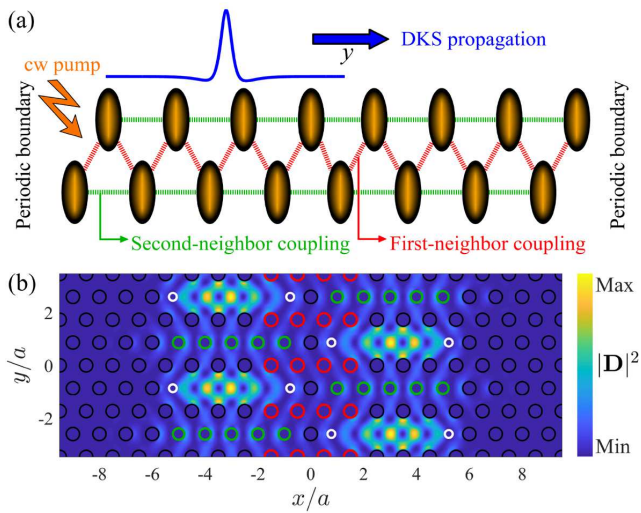}
\caption{{(a) Schematic representation of a CCW system with first and second neighbor cavity coupling. The system is pumped with a cw source and a DKS propagates along the waveguide direction $y$.} (b) Photonic crystal CCW formed by coupled L3 cavities where the red and green holes are allowed to vary in size to tune the first and second neighbor coupling, respectively, while the out-of-plane losses of the waveguide Bloch modes is optimized by varying the position and size of the white holes. The intensity profile of the displacement field [$\mathbf{D}(\mathbf{r})=\epsilon(\mathbf{r})\vec{{\cal E}}(\mathbf{r})$] is shown at the boundary of the Brillouin zone.}\label{fig3}
\end{figure}

In this {work}, we propose a coupled-cavity waveguide (CCW, also known as coupled-resonator optical waveguide or CROW) as a new candidate to efficiently generate low-threshold Kerr frequency combs and DKS. Similar to ring resonators, CCWs define a discrete spectrum of propagating modes which may trigger cascaded parametric FWM in the anomalous dispersion regime, and lead to a highly coherent comb if the waveguide dispersion is conveniently optimized. In fact, since their appearance two decades ago \cite{yariv}, CCWs have shown to be extremely flexible for advanced dispersion engineering and enhancement of non-linear phenomena via slow-light \cite{fan,blair,krauss,notomi1,notomi2,momchil1}. {We show in Fig.~\ref{fig3}(a) a schematic representation of the CCW studied in this work. The periodic waveguide is formed by a staggered distribution of cavities where first (dashed red) and second (dashed green) neighbor cavity-cavity coupling is assumed. The system is pumped with a cw source and a DKS is propagating along the waveguide direction $y$.}

\begin{figure*}[t!]
\centering\includegraphics[width=0.97\textwidth]{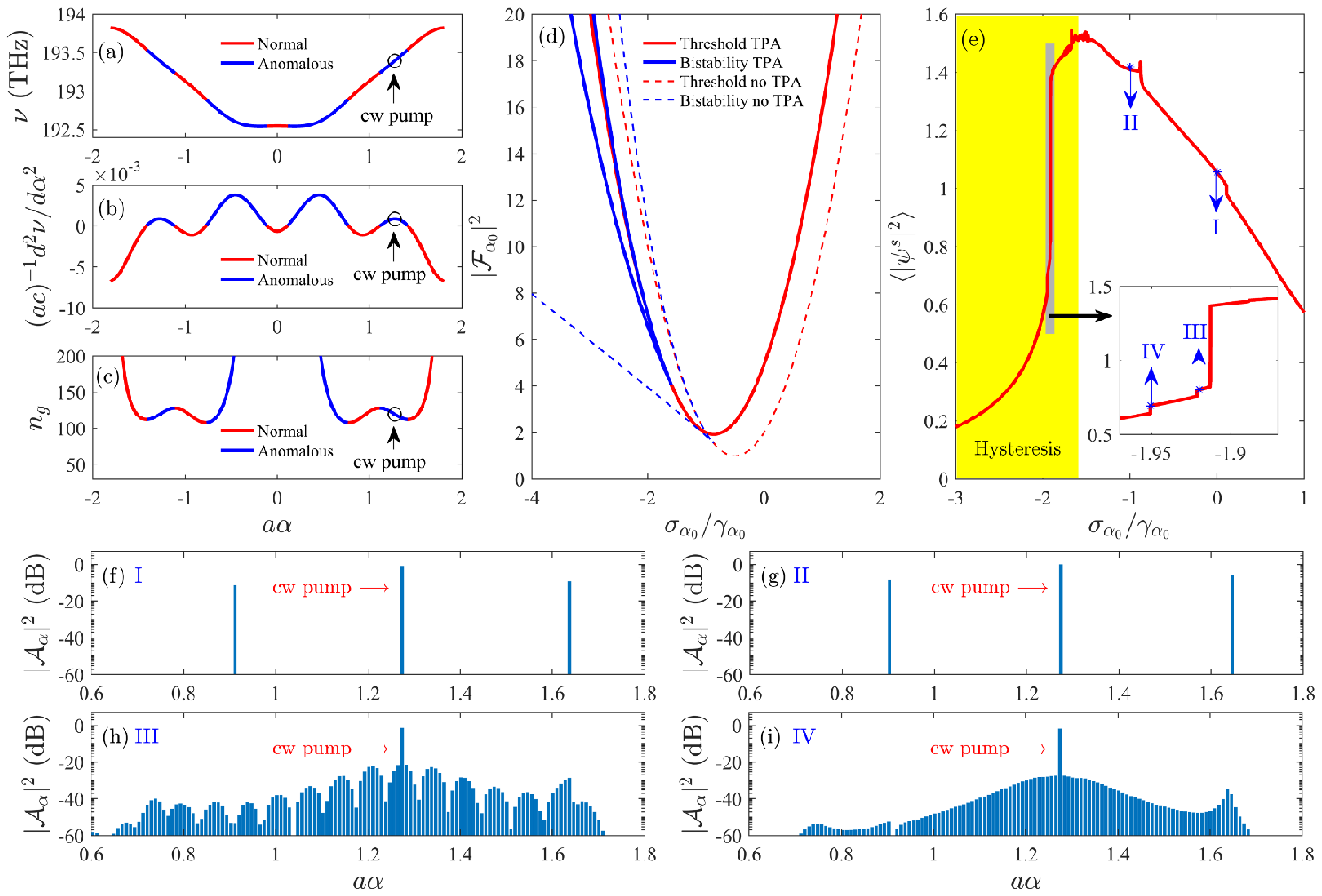}
\caption{(a) Dispersion relation, (b) second order dispersion, and (c) group index in the Brillouin zone of a photonic crystal CCW of 400 L3 cavities. Red and blue curves are where the dispersion is normal and anomalous, respectively. The waveguide is pumped at $a\alpha_0=1.2742$. (d) Threshold (red curve) and bistability boundaries (blue curves) determined by the external pump intensity $|{\cal F}_{\alpha_0}|^2$ as a function of the laser detuning $\sigma_{\alpha_0}$. Dashed curves are for $\beta_{\rm TPA}=0$. (e) Averaged intra-waveguide power of the CCW in the steady state as a function of $\sigma_{\alpha_0}$ for $|{\cal F}_{\alpha_0}|^2=6|{\cal A}_{\alpha_0}|^2_{\rm th}$. Hysteresis arises in the yellow region {$\sigma_{\alpha_0}<-\gamma_{\alpha_0}\sqrt{3}\rho(\kappa)/2$} with $\rho(\kappa=0.2236)=1.84$, and the inset corresponds to a zoom of the rectangular gray area where the discrete steps, signature of switching between soliton states, appear. The corresponding frequency combs at the marked points I-IV in (e) are respectively shown in (f) supercritical Turing pattern of 40-FSR (355.2~GHz) repetition rate, (g) supercritical Turing pattern of 41-FSR (363.3~GHz) repetition rate, (h) soliton molecule of two pulses with single FSR (9.1~GHz) repetition rate, and (i) soliton pulse with single FSR repetition rate. All power quantities are given in {dB units relatively to the} threshold {$|{\cal A}_{\alpha_0}|_{\rm th}^2$}, while detunings are in $\gamma_{\alpha_0}$ units.}\label{fig4}
\end{figure*}

\section{Non-linear photonic crystal CCW}

{The system of Fig.~\ref{fig3}(a) is realized with} the waveguide of coupled L3 photonic crystal slab cavities shown in Fig.~\ref{fig3}(b), where we have plotted the intensity profile of the displacement field [$\mathbf{D}(\mathbf{r})=\epsilon(\mathbf{r})\vec{{\cal E}}(\mathbf{r})$] at the boundary of the Brillouin zone. The photonic crystal is formed by a hexagonal lattice of holes, pitch $a=400$~nm and hole radii $r=0.25a$, etched in a silicon membrane of thickness $d=0.55a$, { and the L3 cavity is introduced by removing three holes along the $\Gamma K$ direction of the lattice}. This CCW configuration allows us to tune first and second neighbor coupling between the L3 cavities by varying the size of the red and green holes, respectively, and optimize the out-of-plane losses by varying the position and size of the white ones. The waveguide of Fig.~\ref{fig3}(b) has been previously proposed as a compact CCW with outstanding figures of merit obtained via automated global optimization \cite{momchil1} and successfully measured experimentally \cite{mohamed,mohamed2}. Here, we adopt the design with largest averaged group index and small second order dispersion reported in Ref.~\citenum{momchil1} with parameters $(\Delta r_1,\Delta r_2, \Delta r_3, \Delta x)=(-0.0049,-0.0340,-0.1016,0.2204)a$, where $r_1=r+\Delta r_1$, $r_2=r+\Delta r_2$ and $r_3=r+\Delta r_3$ are the radii of the red, green and white holes, respectively, and $\Delta x$ is the outward displacement of the latter ones. The dispersion relation $\nu_\alpha$ of the waveguide and decay rates $\gamma_\alpha$ of the Bloch modes of momentum $\alpha$ are calculated with the guided mode expansion method (GME) \cite{gme} for a system length of $M=400$ cavities (400 normal modes within the Brillouin zone of the CCW) with period $l=\sqrt{3}a$ (waveguide length $L=Ml$), while the non-linear mode volume is computed with a commercial FDTD solver \cite{lumerical} and found to be $V_c=0.345~\mu$m$^3$. {This sample length represents a good compromise between total number of normal modes [avoiding finite-size effects in the coupled mode equations Eq.~(\ref{cme7})] and robustness to fabrication disorder \cite{song,mohamed2}.} Silicon parameters at telecom frequencies are considered for the material, namely, dielectric constant $\epsilon=12.04$, Kerr coeficient $n_2=5.52\times10^{-18}$~m$^2/$W and two-photon absorption (TPA) coefficient $\beta_{\rm TPA}=1\times10^{-11}$~m$/$W \cite{jalali}. The GME dispersion relation, second order dispersion and group index are shown in Fig.~\ref{fig4}(a)-\ref{fig4}(c), respectively, for which red and blue segments correspondingly highlight the regions of normal and anomalous dispersion. {Notice that these dispersion curves are valid for a straight waveguide with periodic boundary conditions, nevertheless, we expect them to describe very well a close loop of coupled cavities as bending losses is usually very small, with respect to out-of-plane losses, in photonic crystal geometries \cite{zhang}}. The system is pumped at $a\alpha_0=1.2742$, where $\nu_{\alpha_0}=193.39$~THz, $(ac)^{-1}(d^2\nu/\alpha^2)_{\alpha_0}=8.63\times10^{-4}$, $n_{g,\alpha_0}=119.34$ and the normal mode quality factor is $Q_{\alpha_0}=5.72\times10^4$. This sets an internal mode threshold given by
\begin{equation}\label{Thm}
\left|{\cal A}_{\alpha_0}\right|_{\rm th}^2=\frac{\epsilon V_c}{2ln_{g,\alpha_0}n_2Q_{\alpha_0}}f(\kappa)=138~\mbox{mW},
\end{equation}

\noindent where ${\cal A}_{\alpha_0}$ is the slowly-varying amplitude of the Bloch mode $\alpha_0$, and $f(\kappa)=(\sqrt{1+\kappa^2}+2\kappa)/(1-3\kappa^2)$, with $\kappa=c\beta_{\rm TPA}/(2n_2\omega_{\alpha_0})$, is a function of the material parameters only which gives $f(\kappa=0.2236)=1.73$ for silicon at $\omega_{\alpha_0}/2\pi=\nu_{\alpha_0}=193.39$~THz, and reduces to $f(\kappa=0)=1$ for $\beta_{\rm TPA}=0$. This result is particularly remarkable because, even by considering that TPA increases the threshold by a factor of 1.73, $|{\cal A}_{\alpha_0}|_{\rm th}^2=138~$mW is still one order of magnitude smaller than in typical mm-size crystalline non-linear ring resonators \cite{chembo2}. Equation~(\ref{Thm}) undoubtedly strengths the potential capabilities of CCW structures for low-threshold comb generation. While the factor $V/Q$ also enters in the expression for the internal mode threshold of ring resonators \cite{chembo0,chembo1,chembo3}, CCWs can drastically decrease this minimal power in the slow-light regime, i.e., for $n_{g,\alpha_0} \gg \sqrt{\epsilon}$. Such structural effect allows to access the rich physics of DKS in CCWs at much lower powers than in their monolithic counterparts even in presence of strong TPA, which is the main source of non-linear losses in silicon structures at telecom wavelengths \cite{gaeta4,agrawal}. It is worth emphasizing that, since we are not specifying any particular coupling architecture with the external pump, we are referring to the internal mode threshold which scales as $1/Q$, and not to the external power threshold which is found to scale as $1/Q^2$ for ring resonators at critical coupling with the external source \cite{kippenberg3}. {While the latter is an important parameter to set the excitation power in an experimental setup, the former quantity, also called the minimum intracavity power for comb generation \cite{chembo2}, gives the corresponding effective power into the specific normal mode of the system.} We plot in Fig.~\ref{fig4}(d) the driven intensity $|{\cal F}_{\alpha_0}|^2$, in $|{\cal A}_{\alpha_0}|_{\rm th}^2$ units, evaluated at threshold (continuous red lines) and at the bistability boundaries (continuous blue curves), as a function of the laser detuning $\sigma_{\alpha_0}=\Omega_0-\omega_{\alpha_0}$ in units of $\gamma_{\alpha_0}$, where the dashed lines are the corresponding curves for $\beta_{\rm TPA}=0$. The effects of TPA on the threshold are clearly seen in this plot, moreover, the hysteresis region (where the system has bistable states) is red-shifted and the area between the bistability boundaries is significantly reduced. Hence, the system needs to be pumped stronger and the laser frequency has to be further decreased to see Kerr frequency combs and DKS in presence of TPA. 

\begin{figure}[t!]
\centering\includegraphics[width=0.47\textwidth]{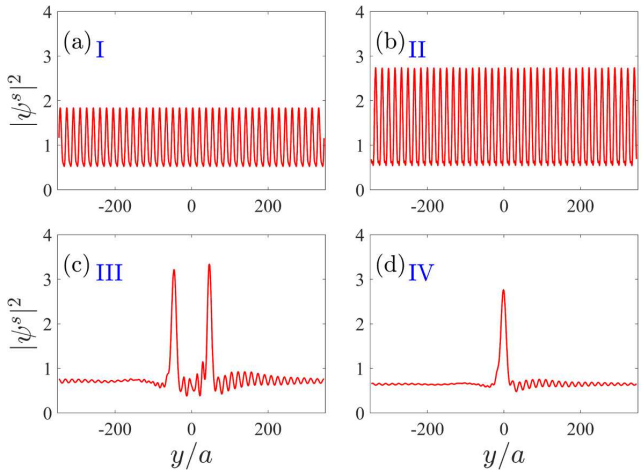}
\caption{Steady state envelope functions at the marked points I-IV in Fig.~\ref{fig4}(e). (a) Supercritical Turing pattern with 40 rolls. (b) Supercritical Turing pattern with 41 rolls. (c) Soliton molecule of two pulses. (d) Single soliton pulse. {$|\psi^s|^2$ is given in units of the power threshold $|{\cal A}_{\alpha_0}|_{\rm th}^2$.}}\label{fig5}. 
\end{figure}

\section{Frequency comb and DKS solutions}

In order to find the possible steady state solutions determined by the non-linear CCW dynamics, we carry out a frequency scan at $|{\cal F}_{\alpha_0}|^2=6|{\cal A}_{\alpha_0}|_{\rm th}^2$ in Fig.~\ref{fig4}(d) within the $\sigma_{\alpha_0}/\gamma_{\alpha_0}$ interval $[-3,1]$. The dynamical non-linear coupled-mode equations Eq.~(\ref{cme7}) are thus propagated in time for each $\sigma_{\alpha_0}$ value along this trajectory using an explicit Runge-Kutta integrator and fast Fourier transform (see Ref.~\citenum{hansson}) until the steady state is reached. Our simulation takes into account the momentum-dependent decay rate $\gamma_{\alpha}$ of the Bloch modes, and the initial condition of the integrator is a sharp Gaussian pulse $\psi(y,0)=\exp[-0.5(y/l)^2]$, where $\psi(y,t)=\sum_\alpha{\cal A}_\alpha(t)e^{i\sigma_\alpha t}e^{-i(\alpha-\alpha_0)y}$ is the envelope function along the CCW direction. Results of this analysis are presented in Fig.~\ref{fig4}(e) where we show the averaged waveguide power in the steady state $\langle|\psi^s|^2\rangle=\int_L|\psi^s(y)|^2dy$, in units of threshold, as a function of the laser detuning. When the laser frequency is red-shifted the intra-waveguide power start to increase until a clear series of discrete steps (see inset), decreasing the average power, are seen in the yellow area, which corresponds to the region where there is hysteresis in the system and the soliton solutions are expected to appear \cite{kippenberg4,chembo4}. In fact, these steps have been previously measured experimentally in the transmission of non-linear ring resonators, and they are associated to the formation of different soliton states within the system, in excellent agreement with the predictions of the non-linear coupled mode equations and the Lugiato-Lefever formalism \cite{kippenberg1}. We show in Figs.~\ref{fig4}(f)-\ref{fig4}(i) the respective frequency combs at the four representative points marked in Fig.~\ref{fig4}(e). Figures~\ref{fig4}(f) and \ref{fig4}(g) displays a sequence of primary combs separated by 40-FSR and 41-FSR, respectively, which correspond to supercritical Turing patterns as they are excited above threshold. In Figs.~\ref{fig4}(h)-\ref{fig4}(i), both combs are subcritical (excited right below threshold) with a single FSR spacing, and they are the signature of soliton complexes. Notably, because of the presence of non-trivial high order terms in the dispersion relation of the photonic crystal CCW, these frequency combs {have} signatures of dispersive wave formation or soliton Cherenkov radiation {at the linear phase matching condition $\omega_\alpha-[\omega_{\alpha_0}+\zeta_1(\alpha-\alpha_0)]=0$, where $\zeta_1$ represents the group velocity \cite{kippenberg2}. Particularly, the dispersive peak clossest to the zero comb line is predicted around $a\alpha\simeq1.63$, which is in perfect agreement with our numerical simulations}. The corresponding steady state intensity of the envelope functions are shown in Fig.~\ref{fig5}. For the I (40 FSR repetition rate) and II (41 FSR repetition rate) states, we see 40 and 41 Turing rolls, respectively, in agreement with the Lugiato-Lefever theory of Kerr frequency combs in the anomalous dispersion regime \cite{chembo4}. Moreover, for the III and IV states we clearly identify, respectively, a soliton molecule composed of two pulses and a single soliton pulse, propagating along the waveguide direction while keeping their shape and amplitude. It must be said that because these soliton structures are subcritical (pumped below threshold), they are extremely sensitive to the initial conditions, and therefore, the system may follow different trajectories in Fig~\ref{fig4}(e) when slightly changing the Bloch mode amplitudes at $t=0$ \cite{chembo4,kippenberg1}. {Furthermore, the characteristic width of the soliton pulse is much smaller than the waveguide length thus ruling out finite-size effects in the basis expansion.} {While results presented in Fig.~\ref{fig4} are strictly 
valid for a CCW system with periodic boundary conditions, we still expect highly coherent combs in finite waveguides where non-negligible corrections to the non-linear interaction are predicted by the theory of frequency combs in Fabry-Perot resonators \cite{lugiato}. Nevertheless, a rigorous quantitative analysis is out of the scope of this paper, and will be focus of a future work.}  

\begin{figure}[t!]
\centering\includegraphics[width=0.47\textwidth]{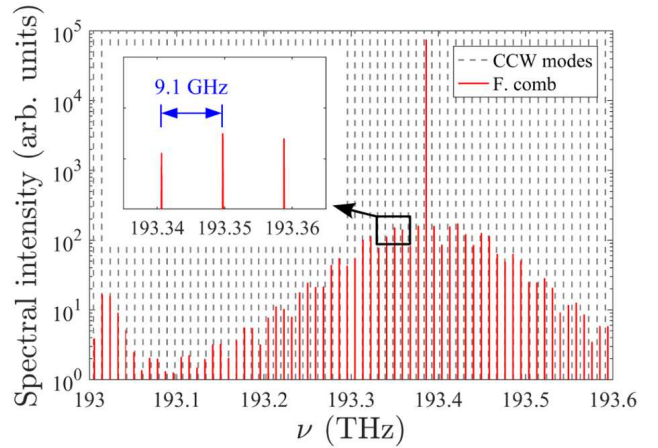}
\caption{Frequency spectrum of the DKS in Fig.~\ref{fig5}(d), where the dashed vertical lines represent the normal modes of the photonic crystal CCW. A small repetition rate of 9.1~GHz is obtained.}\label{fig6}
\end{figure}

Figure~\ref{fig6} shows the frequency spectrum of the DKS in Fig.~\ref{fig5}(d), where the unloaded modes (normal modes of the CCW) are represented by the vertical dashed lines. As expected from energy conservation, the spectral lines are equally spaced with a repetition rate of 9.1~GHz. Such rates have been previously achieved in non-linear ring resonators of 7 mm diameter \cite{vahala}, in contrast to our system length of $\sim277~\mu$m. { Notice that the minimal external power for comb generation may significantly vary depending on the material, geometry and more importantly, the coupling architecture with a bus waveguide \cite{chembo2}, however,} this important result notably highlights the potential capabilities of $\mu$m-scale low-threshold CCW systems to generate low-rate frequency combs, which are desirable in high-precision comb applications \cite{kippenberg4}. 

{It is important to stress that the specific choice of photonic crystal CCWs is mainly motivated by their characteristic diffraction-limited mode volumes \cite{hughes1} which play an important role on decreasing the threshold power and final device size. Nevertheless, the results presented in this section only rely on the dispersion relation of the CCW and are therefore expected for any CCW system, e.g. coupled microrings, displaying anomalous dispersion with, more generally, $n$-th neighbor coupling.} 

\section{Conclusions}

In conclusion, we have presented results on Kerr frequency combs and DKS in CCWs where the main source of non-linear loses is given by TPA. We found that the internal mode threshold for comb generation depends on the main CCW figures of merit, namely, cavity volume, normal mode quality factor and group index. While TPA losses has the main effect of increasing the threshold power and inducing a red shift in the optimal laser detuning, structural slow-light plays an important role on reducing the minimal power to trigger FWM phenomena between the CCW normal modes, evidencing the capabilities of CCW systems for low-threshold frequency combs. Specifically, we have demonstrated the possibility of DKS in a realistic dispersion-engineered silicon photonic crystal CCW at telecom wavelengths, where highly-complex combs were seen with signatures of soliton Cherenkov radiation. Repetition rates of few gigahertz in a $\sim277~\mu$m-length photonic crystal CCW have been obtained into the soliton regime, which are commonly achieved in $\sim7$~mm size ring resonators, thus demonstrating the potential of CCW systems for high-precision comb applications in ultra-compact devices. Finally, although the spectral spanning of the combs generated in CCW is limited to the waveguide bandwidth, our results open the way to Kerr frequency comb and DKS generation via advanced dispersion engineering and slow-light non-linear enhancement.

\onecolumngrid

\appendix

\section{Electromagnetic wave equation}
We start from Maxwell's curl equations
\begin{align}
 \nabla\times\mathbf{E}(\mathbf{r},t)+\mu_0\frac{\partial\mathbf{H}(\mathbf{r},t)}{\partial t} & = 0 \label{curl1}\\
 \nabla\times\mathbf{H}(\mathbf{r},t)-\varepsilon_0\epsilon(\mathbf{r},|\mathbf{E}|^2)\frac{\partial\mathbf{E}(\mathbf{r},t)}{\partial t} &=0,\label{curl2}
\end{align}
where the material is assumed to be non-dispersive, non-linear and isotropic. If we apply the curl operator to Eq.~\ref{curl1} and then use Eq.~\ref{curl2} we get the wave equation for $\mathbf{\mathbf{E}(\mathbf{r},t)}$
\begin{equation}\label{waveE}
\left[\hat{\mathcal{L}}+\frac{\epsilon(\mathbf{r},|\mathbf{E}|^2)}{c^2}\frac{\partial^2}{\partial t^2} \right]\mathbf{E}(\mathbf{r},t)=0,
\end{equation}
where $\hat{\mathcal{L}}=\nabla\times\nabla\times$ and $c=1/\sqrt{\varepsilon_0\mu_0}$.

\section{Electric field expansion}
Because the absolute intensity of the electric field determines the effective strength of the Kerr non-linearity, it is very important to write $\mathbf{E}(\mathbf{r},t)$ in proper units. We specifically adopt the SI system, where the electric field has units of [V/m], and expand it in terms of the of the normal modes of the coupled-cavities waveguide (CCW):
\begin{equation}\label{expE1}
\mathbf{E}(\mathbf{r},t)=\sum_{\mu}{\cal A}^{'}_{\mu}(t)e^{-i\omega_{\mu}t}\mathbf{E}_{\mu}(\mathbf{r})+E_{\rm ext}e^{-i\Omega_0t}\hat{e}_0,
\end{equation}
where ${\cal A}^{'}_{\mu}(t)$ is the slowly-varying amplitude of the normal mode $\mathbf{E}_{\mu}(\mathbf{r})$ with eigenfrequency $\omega_{\mu}$ and momentum $\mu$ (along the waveguide direction $\hat{\mu}$), and $E_{\rm ext}$ is the electric field amplitude of the driving continuous-wave (cw) source with frequency $\Omega_0$ and vector direction $\hat{e}_0$. Equation \ref{expE1} is the so called slowly-varying amplitude expansion, as the time and spatial variations have been separated, and the fast-varying temporal dependence is explicitly represented by the oscillatory term $e^{-i\omega_{\mu}t}$. Notice that the field $\mathbf{E}_{\mu}(\mathbf{r})$ in Eq. \ref{expE1} also has units of [V/m], then ${\cal A}^{'}$ is dimensionless. In order to express the CCW normal modes in a more convenient form we write
\begin{equation}\label{Emu}
 \mathbf{E}_{\mu}(\mathbf{r})=\sqrt{N_{\mu}}\vec{{\cal E}}_{\mu}(\mathbf{r}),
\end{equation}
where $N_{\mu}$ is a normalization factor and $\vec{{\cal E}}_{\mu}(\mathbf{r})$ fulfills the generalized eigenvalue equation
\begin{equation}\label{eigenE}
 \hat{\mathcal{L}}\vec{{\cal E}}_{\mu}(\mathbf{r})=\frac{\tilde{\omega}_\mu^2}{c^2}\epsilon(\mathbf{r})\vec{{\cal E}}_{\mu}(\mathbf{r}),
\end{equation}
subject to the normalization condition
\begin{equation}\label{ortho}
\int_{l} \epsilon(\mathbf{r})\vec{{\cal E}}^\ast_{\alpha}(\mathbf{r})\cdot\vec{{\cal E}}_{\mu}(\mathbf{r})dV=\delta_{\alpha,\mu},
\end{equation}
with $l$ representing the period of the waveguide and $\epsilon(\mathbf{r})$ its dielectric function. Equation~\ref{ortho} sets the units of $\vec{{\cal E}}_{\mu}(\mathbf{r})$ to [m$^{-3/2}$]. In Eq.~\ref{eigenE} we have considered the complex frequency 
\begin{equation}\label{complexw}
 \tilde{\omega}_\mu=\omega_\mu-i\frac{\gamma_\mu}{2},
\end{equation}
where $\gamma_\mu$ is the overall loss rate of the Bloch mode $\vec{{\cal E}}_{\mu}(\mathbf{r})$. For the particular case of a photonic crystal slab, $\gamma_\mu$ represents the out-of-plane losses due to the coupling with the leaky modes of the homogeneous system \cite{gme}. Notice that the normalization condition of Eq.~\ref{ortho}, satisfied by normal modes, is not rigorously valid when the eigenvalue of Eq.~\ref{eigenE} is complex, i.e., when dealing with quasinormal modes \cite{hughes1}. Nevertheless it still represents a very good approximation in the low loss limit $\gamma_\mu\ll\omega_\mu$, which is the case of interest in the present formulation. The normalization factor $N_\mu$ can be easily found by recalling the fundamental relation between the group velocity $v_g(\mu)$, averaged energy flux and averaged energy density of the electromagnetic field, which holds in periodic waveguides \cite{joannopoulos}
\begin{equation}\label{vg1}
 v_g(\mu)=\frac{\int_l \hat{\mu}\cdot\mathbf{S}_\mu dV}{{\cal U}_{\mathbf{E}_\mu}+{\cal U}_{\mathbf{H}_\mu}},
\end{equation}
where $\mathbf{S}_\mu(\mathbf{r})=\mbox{Re}[\mathbf{E}^\ast_{\mu}(\mathbf{r})\times\mathbf{H}^\ast_{\mu}(\mathbf{r})]/2$ is the time-averaged Poynting vector (energy flux), ${\cal U}_{\mathbf{E}_\mu}=\int_l\varepsilon_0\epsilon(\mathbf{r})|\mathbf{E}_{\mu}(\mathbf{r})|^2dV/4$ is the physical electric field energy (integrated density energy) and ${\cal U}_{\mathbf{H}_\mu}=\int_l\mu_0|\mathbf{H}_{\mu}(\mathbf{r})|^2dV/4$ the magnetic one in the waveguide period. If we set the waveguide direction along the $y$ axis, i.e., $\hat{\mu}=\hat{y}$, and use the result that for harmonic modes the electric and magnetic field energies are equal \cite{joannopoulos}, i.e., ${\cal U}_{\mathbf{E}_\mu}={\cal U}_{\mathbf{H}_\mu}$, Eq. \ref{vg1} takes the following form
\begin{equation}\label{vg2}
 v_g(\mu)=\frac{\int_ldy\left[\int_\infty dxdz S_{\mu,y}\right]}{2{\cal U}_{\mathbf{E}_\mu}},
\end{equation}
where the last integral is carried out over all space and it is the power flux $P_\mu=\int_\infty dxdz S_{\mu,y}$ through the transverse cross section of the waveguide, which has been shown to be independent of $y$ \cite{song}, therefore
\begin{equation}\label{vg3}
 v_g(\mu)=\frac{lP_\mu}{2{\cal U}_{\mathbf{E}_\mu}}.
\end{equation}
Using the definition of ${\cal U}_{\mathbf{E}_\mu}$, Eq. \ref{vg3} turns into
\begin{equation}\label{vg4}
 v_g(\mu)=\frac{lP_\mu}{\frac{1}{2}\int_l\varepsilon_0\epsilon(\mathbf{r})|\mathbf{E}_{\mu}(\mathbf{r})|^2dV},
\end{equation}
which means that the electric field $\mathbf{E}_{\mu}(\mathbf{r})$ must fulfill
\begin{equation}\label{enorm}
 \int_l\epsilon(\mathbf{r})|\mathbf{E}_{\mu}(\mathbf{r})|^2dV=\frac{2lP_\mu}{\varepsilon_0v_g(\mu)}.
\end{equation}
The expression for $N_\mu$ is readily found from Eqs.~\ref{Emu}, \ref{ortho} and \ref{enorm}
\begin{equation}\label{norm}
 N_\mu=\frac{2lP_\mu}{\varepsilon_0v_g(\mu)},
\end{equation}
and the field expansion in Eq. \ref{expE1} is thus expressed in terms of the electric field eigenmodes $\vec{{\cal E}}_{\mu}(\mathbf{r})$ as follows
\begin{equation}\label{expE2}
 \mathbf{E}(\mathbf{r},t)=\sqrt{\frac{2l}{\varepsilon_0}}\sum_{\mu}\sqrt{P_\mu}{\cal A}^{'}_{\mu}(t)e^{-i\omega_{\mu}t}\frac{1}{\sqrt{v_g(\mu)}}\vec{{\cal E}}_{\mu}(\mathbf{r})+E_{\rm ext}e^{-i\Omega_0t}\hat{e}_0.
\end{equation}
We now define the new amplitude ${\cal A}_{\mu}(t)=\sqrt{P_\mu}{\cal A}^{'}_{\mu}(t)$ and simplify the notation for the group velocity of the mode $\mu$ as $v_g(\mu) \rightarrow v_\mu$. The electric field expansion takes the following final form
\begin{equation}\label{expEf}
  \mathbf{E}(\mathbf{r},t)=\sqrt{\frac{2l}{\varepsilon_0}}\sum_{\mu}{\cal A}_{\mu}(t)e^{-i\omega_{\mu}t}\frac{1}{\sqrt{v_\mu}}\vec{{\cal E}}_{\mu}(\mathbf{r})+E_{\rm ext}e^{-i\Omega_0t}\hat{e}_0,
\end{equation}
where $|{\cal A}_{\mu}(t)|^2$ is the instantaneous power, in [W] units, of the mode $\mu$ propagating in the waveguide direction.

\section{Non-linear coupled mode equations}\label{cmesec}
Let's now consider a CCW of $M$ cavities where the total electric field of the system is represented by Eq.~\ref{expEf} and the normalization condition of Eq.~\ref{ortho} is written as
\begin{equation}\label{orthoM}
\int_{L} \epsilon(\mathbf{r})\vec{{\cal E}}^\ast_{\alpha}(\mathbf{r})\cdot\vec{{\cal E}}_{\mu}(\mathbf{r})dV=M\delta_{\alpha,\mu},
\end{equation}
when integrating over the total waveguide length $L=Ml$. The second time-derivative of Eq.~\ref{expEf} reads
\begin{equation}\label{secder}
\frac{\partial^2\mathbf{E}(\mathbf{r},t)}{\partial t^2}=\sqrt{\frac{2l}{\varepsilon_0}}\sum_{\mu}\left[\ddot{{\cal A}}_{\mu}(t)-2i\omega_\mu\dot{{\cal A}}_{\mu}(t)-\omega_\mu^2{\cal A}_{\mu}(t)\right]e^{-i\omega_{\mu}t}\frac{1}{\sqrt{v_\mu}}\vec{{\cal E}}_{\mu}(\mathbf{r})-\Omega_0^2E_{\rm ext}e^{-i\Omega_0t}\hat{e}_0,
\end{equation}
while the operator $\hat{\mathcal{L}}$ applied to  Eq.~\ref{expEf} gives
\begin{equation}\label{opterm}
 \hat{\mathcal{L}}\mathbf{E}(\mathbf{r},t)=\sqrt{\frac{2l}{\varepsilon_0}}\frac{1}{c^2}\sum_{\mu}\tilde{\omega}_\mu^2{\cal A}_{\mu}(t)e^{-i\omega_{\mu}t}\frac{1}{\sqrt{v_\mu}}\epsilon(\mathbf{r})\vec{{\cal E}}_{\mu}(\mathbf{r}),
\end{equation}
where we have used Eq.~\ref{eigenE}. Because $\mathbf{E}(\mathbf{r},t)$ must satisfy the wave equation in Eq.~\ref{waveE}, we employ Eqs.~\ref{secder} and \ref{opterm} to get
\begin{align}\label{waveEexp1}
 &\sum_{\mu}\tilde{\omega}_\mu^2{\cal A}_{\mu}(t)e^{-i\omega_{\mu}t}\frac{1}{\sqrt{v_\mu}}\epsilon(\mathbf{r})\vec{{\cal E}}_{\mu}(\mathbf{r}) \nonumber \\ 
 &+ \epsilon(\mathbf{r},|\mathbf{E}|^2) \left\{ \sum_{\mu}\left[\ddot{{\cal A}}_{\mu}(t)-2i\omega_\mu\dot{{\cal A}}_{\mu}(t)-\omega_\mu^2{\cal A}_{\mu}(t)\right]e^{-i\omega_{\mu}t}\frac{1}{\sqrt{v_\mu}}\vec{{\cal E}}_{\mu}(\mathbf{r}) - \sqrt{\frac{\varepsilon_0}{2l}}\Omega_0^2E_{\rm ext}e^{-i\Omega_0t}\hat{e}_0 \right\}=0.
\end{align}
The intensity dependent dielectric function in Eq.~\ref{waveEexp1} can be set to the reference value $\epsilon(\mathbf{r})=\epsilon(\mathbf{r},0)$ in the pump contribution as the perturbation induced by $|\mathbf{E}|^2$ can be neglected when driving the system at $\epsilon(\mathbf{r})$. Therefore, Eq.~\ref{waveEexp1} turns into
\begin{align}\label{waveEexp2}
&\sum_{\mu}\tilde{\omega}_\mu^2{\cal A}_{\mu}(t)e^{-i\omega_{\mu}t}\frac{1}{\sqrt{v_\mu}}\epsilon(\mathbf{r})\vec{{\cal E}}_{\mu}(\mathbf{r}) \nonumber \\
&+\epsilon(\mathbf{r},|\mathbf{E}|^2)\sum_{\mu}\left[\ddot{{\cal A}}_{\mu}(t)-2i\omega_\mu\dot{{\cal A}}_{\mu}(t)-\omega_\mu^2{\cal A}_{\mu}(t)\right]e^{-i\omega_{\mu}t}\frac{1}{\sqrt{v_\mu}}\vec{{\cal E}}_{\mu}(\mathbf{r}) - \epsilon(\mathbf{r})\sqrt{\frac{\varepsilon_0}{2l}}\Omega_0^2E_{\rm ext}e^{-i\Omega_0t}\hat{e}_0 =0.
\end{align}
At this point we explicitly write the intensity dependent dielectric function as \cite{boyd}
\begin{equation}\label{eps1}
\epsilon(\mathbf{r},|\mathbf{E}|^2)=\left[ n(\mathbf{r})+n_2(\mathbf{r})I(\mathbf{r},|\mathbf{E}|^2)+i\frac{c}{2\omega_0}\beta_{\rm TPA}(\mathbf{r})I(\mathbf{r},|\mathbf{E}|^2) \right]^2,
\end{equation}
where $n(\mathbf{r})=\sqrt{\epsilon(\mathbf{r})}$ is the linear refractive index, $n_2(\mathbf{r})$ and $\beta_{\rm TPA}(\mathbf{r})$ are the spatial dependent Kerr and two-photon absorption (TPA) coefficients, respectively, of the material at $\omega_0$  and 
\begin{equation}\label{Intensity}
I(\mathbf{r},|\mathbf{E}|^2)=\frac{\varepsilon_0c}{2}n(\mathbf{r})|\mathbf{E}|^2,
\end{equation}
is the instantaneous electric field intensity. As the non-linear shift induced on the dielectric function is expected to be small, Eq.~\ref{eps1} can be approximated to
\begin{equation}\label{eps2}
 \epsilon(\mathbf{r},|\mathbf{E}|^2)\backsimeq \epsilon(\mathbf{r})+2n(\mathbf{r})\Delta n(\mathbf{r},|\mathbf{E}|^2),
\end{equation}
with
\begin{equation}\label{Dn}
 \Delta n(\mathbf{r},|\mathbf{E}|^2)=\frac{\varepsilon_0c}{2}n(\mathbf{r})\left[ n_2(\mathbf{r})+i\frac{c}{2\omega_0}\beta_{\rm TPA}(\mathbf{r}) \right]|\mathbf{E}|^2.
\end{equation}
Replacing Eq.~\ref{eps2} in Eq.~\ref{waveEexp2} gives (after reorganizing terms)
\begin{align}\label{waveEexp3}
 & \sum_{\mu}(\tilde{\omega}_\mu^2-\omega_\mu^2){\cal A}_{\mu}(t)e^{-i\omega_{\mu}t}\frac{1}{\sqrt{v_\mu}}\epsilon(\mathbf{r})\vec{{\cal E}}_{\mu}(\mathbf{r}) + \sum_{\mu}\left[\ddot{{\cal A}}_{\mu}(t)-2i\omega_\mu\dot{{\cal A}}_{\mu}(t)\right]e^{-i\omega_{\mu}t}\frac{1}{\sqrt{v_\mu}}\epsilon(\mathbf{r})\vec{{\cal E}}_{\mu}(\mathbf{r}) \nonumber \\
 & -2\sum_{\mu}\omega_\mu^2{\cal A}_{\mu}(t)e^{-i\omega_{\mu}t}\frac{1}{\sqrt{v_\mu}}n(\mathbf{r})\Delta n(\mathbf{r},|\mathbf{E}|^2)\vec{{\cal E}}_{\mu}(\mathbf{r}) \nonumber \\
 & + 2\sum_{\mu}\left[\ddot{{\cal A}}_{\mu}(t)-2i\omega_\mu\dot{{\cal A}}_{\mu}(t)\right]e^{-i\omega_{\mu}t}\frac{1}{\sqrt{v_\mu}}n(\mathbf{r})\Delta n(\mathbf{r},|\mathbf{E}|^2)\vec{{\cal E}}_{\mu}(\mathbf{r}) - \epsilon(\mathbf{r})\sqrt{\frac{\varepsilon_0}{2l}}\Omega_0^2E_{\rm ext}e^{-i\Omega_0t}\hat{e}_0 =0.
\end{align}
In order to arrive to a system of differential equations for the slowly-varying amplitudes, Eq.~\ref{waveEexp3} is multiplied by $\vec{{\cal E}}^\ast_{\alpha}(\mathbf{r})$ and integrated over the whole waveguide. After the orthogonality condition of Eq.~\ref{orthoM} is explicitly applied, we get 
\begin{align}\label{cme1}
 & (\tilde{\omega}_\alpha^2-\omega_\alpha^2){\cal A}_{\alpha}(t)e^{-i\omega_{\alpha}t}\frac{1}{\sqrt{v_\alpha}} + \left[\ddot{{\cal A}}_{\alpha}(t)-2i\omega_\alpha\dot{{\cal A}}_{\alpha}(t)\right]e^{-i\omega_{\alpha}t}\frac{1}{\sqrt{v_\alpha}} \nonumber \\
 & -\frac{2}{M}\sum_{\mu}\omega_\mu^2{\cal A}_{\mu}(t)e^{-i\omega_{\mu}t}\frac{1}{\sqrt{v_\mu}}\int_{L}n(\mathbf{r})\Delta n(\mathbf{r},|\mathbf{E}|^2)\vec{{\cal E}}^\ast_{\alpha}(\mathbf{r})\cdot\vec{{\cal E}}_{\mu}(\mathbf{r})dV \nonumber \\
 & +\frac{2}{M}\sum_{\mu}\left[\ddot{{\cal A}}_{\mu}(t)-2i\omega_\mu\dot{{\cal A}}_{\mu}(t)\right]e^{-i\omega_{\mu}t}\frac{1}{\sqrt{v_\mu}}\int_{L}n(\mathbf{r})\Delta n(\mathbf{r},|\mathbf{E}|^2)\vec{{\cal E}}^\ast_{\alpha}(\mathbf{r})\cdot\vec{{\cal E}}_{\mu}(\mathbf{r})dV \nonumber \\
 & -\sqrt{\frac{\varepsilon_0}{2l}}\Omega_0^2E_{\rm ext}e^{-i\Omega_0t}\int_{l}\epsilon(\mathbf{r})\vec{{\cal E}}^\ast_{\alpha}(\mathbf{r})\cdot\hat{e}_0dV = 0.
\end{align}
We now apply the slowly-varying assumption where the second derivative terms are assumed to be small, i.e., $|\ddot{{\cal A}}_{\alpha}(t)|\ll |2\omega_\alpha\dot{{\cal A}}_{\alpha}(t)|$, and consequently they can be neglected, as well as the first derivative term multiplying the non-linear perturbation $\Delta n(\mathbf{r},|\mathbf{E}|^2)$ in the second integral of Eq.~\ref{cme1}. We also assume low-loss CCWs, for which we can approximate $\tilde{\omega}_\alpha^2-\omega_\alpha^2\backsimeq-i\omega_\alpha\gamma_\alpha$ in the first term of the equation. Therefore, Eq.~\ref{cme1} is rewritten as
\begin{align}\label{cme2}
\dot{{\cal A}}_{\alpha}(t)e^{-i\omega_{\alpha}t} + \frac{\gamma_\alpha}{2}{\cal A}_{\alpha}(t)e^{-i\omega_{\alpha}t}-\frac{i\sqrt{v_\alpha}}{\omega_\alpha M}\sum_{\mu}\omega_\mu^2{\cal A}_{\mu}(t)e^{-i\omega_{\mu}t}\frac{1}{\sqrt{v_\mu}}\int_{L}n(\mathbf{r})\Delta n(\mathbf{r},|\mathbf{E}|^2)\vec{{\cal E}}^\ast_{\alpha}(\mathbf{r})\cdot\vec{{\cal E}}_{\mu}(\mathbf{r})dV
 -\frac{\gamma_\alpha}{2}{\cal F}_\alpha e^{-i\Omega_0t} = 0
\end{align}
where we have defined the driving amplitude
\begin{equation}\label{Fpump}
 {\cal F}_\alpha=\frac{i\Omega_0^2}{\omega_\alpha\gamma_\alpha}\sqrt{\frac{\varepsilon_0v_\alpha}{2l}}E_{\rm ext}\int_{l}\epsilon(\mathbf{r})\vec{{\cal E}}^\ast_{\alpha}(\mathbf{r})\cdot\hat{e}_0dV.
\end{equation}
Let's focus on the non-linear term of Eq.~\ref{cme2}. From the field expansion in Eq.~\ref{expEf}, the intra-waveguide modulus squared field is given by
\begin{equation}\label{E2}
 |\mathbf{E}(\mathbf{r},t)|^2=\frac{2l}{\varepsilon_0}\sum_{\eta\xi}{\cal A}_{\eta}^\ast(t)e^{i\omega_\eta t}{\cal A}_\xi(t)e^{-i\omega_\xi t}\frac{1}{\sqrt{v_\eta v_\xi}}\vec{{\cal E}}^\ast_{\eta}(\mathbf{r})\cdot\vec{{\cal E}}_{\xi}(\mathbf{r}),
\end{equation}
therefore, the non-linear perturbation in Eq.~\ref{Dn} is written as 
\begin{equation}\label{DnE2}
 \Delta n(\mathbf{r},|\mathbf{E}|^2)=lc\sum_{\eta\xi}{\cal A}_{\eta}^\ast(t)e^{i\omega_\eta t}{\cal A}_\xi(t)e^{-i\omega_\xi t}\frac{1}{\sqrt{v_\eta v_\xi}}n(\mathbf{r})\left[ n_2(\mathbf{r})+i\frac{c}{2\omega_0}\beta_{\rm TPA}(\mathbf{r}) \right]\vec{{\cal E}}^\ast_{\eta}(\mathbf{r})\cdot\vec{{\cal E}}_{\xi}(\mathbf{r}),
\end{equation}
and the integral of Eq.~\ref{cme2} becomes
\begin{align}\label{Nint1}
  \int_{L}n(\mathbf{r})\Delta n(\mathbf{r},|\mathbf{E}|^2)\vec{{\cal E}}^\ast_{\alpha}(\mathbf{r})\cdot\vec{{\cal E}}_{\mu}(\mathbf{r})dV= 
 lc\sum_{\eta\xi}{\cal A}_{\eta}^\ast(t)e^{i\omega_\eta t}{\cal A}_\xi(t)e^{-i\omega_\xi t}\frac{1}{\sqrt{v_\eta v_\xi}}\int_{L}\tilde{n}_2(\mathbf{r})\epsilon(\mathbf{r})\left[\vec{{\cal E}}^\ast_{\alpha}(\mathbf{r})\cdot\vec{{\cal E}}_{\mu}(\mathbf{r})\right] \left[\vec{{\cal E}}^\ast_{\eta}(\mathbf{r})\cdot\vec{{\cal E}}_{\xi}(\mathbf{r})\right]dV,
\end{align}
where we have defined the complex Kerr coefficient
\begin{equation}\label{complexn2}
\tilde{n}_2(\mathbf{r})=n_2(\mathbf{r})+i\frac{c}{2\omega_0}\beta_{\rm TPA}(\mathbf{r}).
\end{equation}
In order to write the last integral term of Eq.~\ref{Nint1} in a more convenient form, we assume that the normal modes of the CCW can be expanded in the cavity mode basis
\begin{equation}\label{cavityexp}
 \vec{{\cal E}}_{\mu}(\mathbf{r})=\sum_mC_m(\mu)\vec{\Upsilon}_m(\mathbf{r}),
\end{equation}
which is a very good approximation as long as the modes are strongly confined in the cavity region \footnote{Rigorously, one should employ the set of Wannier functions to carry out the expansion of Eq.~\ref{cavityexp}, as the cavity modes do not define a complete orthogonal set of functions. Nevertheless, for strongly localized cavity modes this expansion describes very well the physics of a CCW system.}. In Eq.~\ref{cavityexp}, $\vec{\Upsilon}_m(\mathbf{r})$ is the eigenmode of the $m$-th cavity in units of [m$^{-3/2}$] and normalized as
\begin{equation}\label{normcavity}
\int_\infty\epsilon_c(\mathbf{r})|\vec{\Upsilon}_m(\mathbf{r})|^2dV=1,
\end{equation}
with $\epsilon_c(\mathbf{r})$ being the dielectric function of the single cavity system. By replacing Eq.~\ref{cavityexp} into Eq.~\ref{Nint1}, the non-linear integral can be rewritten as
\begin{align}\label{Nint2}
  \int_{L}n(\mathbf{r})\Delta n(\mathbf{r},|\mathbf{E}|^2)\vec{{\cal E}}^\ast_{\alpha}(\mathbf{r})\cdot\vec{{\cal E}}_{\mu}(\mathbf{r})dV=
  lc\sum_{\eta\xi}{\cal A}_{\eta}^\ast(t)e^{i\omega_\eta t}{\cal A}_\xi(t)e^{-i\omega_\xi t}\frac{1}{\sqrt{v_\eta v_\xi}}\sum_{lmhp}C_l^\ast(\alpha)C_m(\mu)C_h^\ast(\eta)C_p(\xi)\Pi_{lmhp},
\end{align}
where the non-linear overlapping elements $\Pi_{lmhp}$ are given by
\begin{equation}\label{Noverlap}
 \Pi_{lmhp}=\int_{L}\tilde{n}_2(\mathbf{r})\epsilon(\mathbf{r})\left[\vec{\Upsilon}^\ast_l(\mathbf{r})\cdot\vec{\Upsilon}_m(\mathbf{r})\right] \left[\vec{\Upsilon}^\ast_h(\mathbf{r})\cdot\vec{\Upsilon}_p(\mathbf{r})\right]dV.
\end{equation}
Since the mode $\vec{\Upsilon}_{m}(\mathbf{r})$ is strongly confined in the cavity region, the Kerr shift on cavity $m$ induced by the field of neighbor cavities ($m\pm1$, $m\pm2$, $\cdots$), is expected to be very small in comparison to the shift induced by its own field strength, therefore, we can safely neglect the cross-Kerr terms in Eq.~\ref{Noverlap} and consider only the diagonal one in the sum of Eq.~\ref{Nint2}, i.e., 
\begin{align}\label{Nint3}
  \int_{L}n(\mathbf{r})\Delta n(\mathbf{r},|\mathbf{E}|^2)\vec{{\cal E}}^\ast_{\alpha}(\mathbf{r})\cdot\vec{{\cal E}}_{\mu}
 (\mathbf{r})dV\backsimeq
  lc\sum_{\eta\xi}{\cal A}_{\eta}^\ast(t)e^{i\omega_\eta t}{\cal A}_\xi(t)e^{-i\omega_\xi t}\frac{1}{\sqrt{v_\eta v_\xi}}\sum_{m}C_m^\ast(\alpha)C_m(\mu)C_m^\ast(\eta)C_m(\xi)\Pi_{m},
\end{align}
with 
\begin{equation}\label{cE41}
 \Pi_{m}=\int_{L}\tilde{n}_2(\mathbf{r})\epsilon(\mathbf{r})|\vec{\Upsilon}_{m}(\mathbf{r})|^4dV.
\end{equation}
Furthermore, the coefficient $\tilde{n}_2(\mathbf{r})$ is non-zero only within the non-linear material, which allow us to write Eq.~\ref{cE41} as
\begin{equation}\label{cE42}
 \Pi_{m}=\tilde{n}_2\epsilon\int_{L}\Theta(\mathbf{r})|\vec{\Upsilon}_{m}(\mathbf{r})|^4dV.
\end{equation}
where $\tilde{n}_2$ is the same of Eq.~\ref{complexn2} but with no spatial dependence, $\epsilon$ is the dielectric constant of the non-linear material, and the function $\Theta(\mathbf{r})$ is defined as
\begin{equation}\label{thetadef}
 \Theta(\mathbf{r})=\begin{cases}
                     1 & \text{non-linear material}\\
                     0 & \text{elsewhere}
                    \end{cases}
\end{equation}
In order to write Eq.~\ref{cE42} in a more intuitive way, it is an excellent approximation to express the integral of $|\vec{\Upsilon}_{m}(\mathbf{r})|^4$ in terms of the cavity non-linear mode volume $V_m$ as follows
\begin{equation}\label{nlV}
 V_m=\frac{\left[\int_\infty\epsilon_c(\mathbf{r})|\vec{\Upsilon}_m(\mathbf{r})|^2dV\right]^2}{\int_\infty\epsilon_c^2(\mathbf{r})|\vec{\Upsilon}_{m}(\mathbf{r})|^4dV}=\frac{1}{\int_\infty\epsilon_c^2(\mathbf{r})|\vec{\Upsilon}_{m}(\mathbf{r})|^4dV}\backsimeq\frac{1}{\epsilon^2\int_{L}\Theta(\mathbf{r})|\vec{\Upsilon}_{m}(\mathbf{r})|^4dV}.
\end{equation}
Therefore
\begin{equation}\label{cE43}
 \Pi_{m}\backsimeq\frac{\tilde{n}_2}{\epsilon V_m}.
\end{equation}
Using Eqs.~\ref{Nint3} and \ref{cE43} the system of Eq.~\ref{cme2} can be written as
\begin{align}\label{cme3}
&\dot{{\cal A}}_{\alpha}(t)e^{i\sigma_{\alpha}t} + \frac{\gamma_\alpha}{2}{\cal A}_{\alpha}(t)e^{i\sigma_{\alpha}t} \nonumber \\
&-\frac{i\sqrt{v_\alpha}lc\tilde{n}_2}{\omega_\alpha\epsilon M}\sum_{\mu\eta\xi}\omega_\mu^2{\cal A}_{\mu}(t)e^{i\sigma_{\mu}t}{\cal A}_{\eta}^\ast(t)e^{-i\sigma_\eta t}{\cal A}_\xi(t)e^{i\sigma_\xi t}\frac{1}{\sqrt{v_\mu v_\eta v_\xi}}\sum_{m}\frac{1}{V_m}C_m^\ast(\alpha)C_m(\mu)C_m^\ast(\eta)C_m(\xi) \nonumber\\
&-\frac{\gamma_\alpha}{2}{\cal F}_\alpha\delta_{\alpha,\alpha_0}=0,
\end{align}
where we have introduced the frequency detuning of the mode $\mu$ with respect to the cw frequency $\Omega_0$
\begin{equation}\label{detuningmu}
 \sigma_\mu=\Omega_0-\omega_\mu,
\end{equation}
and explicitly written the resonance condition for the driven term, which excites only the normal mode $\alpha_0$ whose eigenfrequency is closest to the laser frequency $\Omega_0$. The explicit time dependence of Eq.~\ref{cme3} can be removed by carrying out the following transformation
\begin{equation}\label{Btransf}
{\cal B}_\mu(t)={\cal A}_\mu(t)e^{i\sigma_\mu t},
\end{equation}
which turns Eq.~\ref{cme3} into
\begin{align}\label{cme4}
 \dot{{\cal B}}_{\alpha}(t) + \left[\frac{\gamma_\alpha}{2}-i\sigma_\alpha\right]{\cal B}_{\alpha}(t)
 -\frac{i\sqrt{v_\alpha}lc\tilde{n}_2}{\omega_\alpha\epsilon M}\sum_{\mu\eta\xi}\frac{\omega_\mu^2{\cal B}_{\mu}(t){\cal B}_{\eta}^\ast(t){\cal B}_\xi(t)}{\sqrt{v_\mu v_\eta v_\xi}}\sum_{m}\frac{1}{V_m}C_m^\ast(\alpha)C_m(\mu)C_m^\ast(\eta)C_m(\xi)-\frac{\gamma_\alpha}{2}{\cal F}_\alpha\delta_{\alpha,\alpha_0}=0 \nonumber\\
\end{align}
Equation~\ref{cme4} dictates the non-linear dynamics of the slowly-varying normal mode amplitudes in a CCW system with different coupled cavities, and no particular choice on the boundary conditions. This equation can be highly simplified if we consider identical single mode cavities and periodic boundary conditions, in which the expansion coefficients in the cavity basis are analytical \cite{yariv}
\begin{equation}\label{Ccoeff}
 C_m(\mu)=e^{-iml\mu},
\end{equation}
allowing to evaluate the second sum in Eq.~\ref{cme4}
\begin{align}\label{sumC}
 \sum_{m}\frac{1}{V_m}C_m^\ast(\alpha)C_m(\mu)C_m^\ast(\eta)C_m(\xi) =& \frac{1}{V_c}\sum_m e^{-iml(\mu-\alpha+\xi-\eta)} 
 = \frac{1}{V_c}\frac{\sin\left[(\mu-\alpha+\xi-\eta)Ml/2\right]}{\sin\left[(\mu-\alpha+\xi-\eta)l/2\right]} 
 \backsimeq \frac{M}{V_c}\delta_{\xi,\alpha+\eta-\mu}
\end{align}
where $V_c$ represents the cavity non-linear mode volume and the last approximation is valid for large $M$. Physically, the $\delta$ term in Eq.~\ref{sumC} represents the momentum conservation which is naturally expected from the periodic boundary condition assumption. Using this result, Eq.~\ref{cme4} takes the following form
\begin{equation}\label{cme5}
 \dot{{\cal B}}_{\alpha}(t) + \left[\frac{\gamma_\alpha}{2}-i\sigma_\alpha\right]{\cal B}_{\alpha}(t)-\frac{i\sqrt{v_\alpha}lc\tilde{n}_2}{\omega_\alpha\epsilon V_c}\sum_{\mu\eta}\frac{\omega_\mu^2{\cal B}_{\mu}(t){\cal B}_{\eta}^\ast(t){\cal B}_{\alpha+\eta-\mu}(t)}{\sqrt{v_\mu v_\eta v_{\alpha+\eta-\mu}}}-\frac{\gamma_\alpha}{2}{\cal F}_\alpha\delta_{\alpha,\alpha_0}=0.
\end{equation}
Equation~\ref{cme5} can be further simplified by assuming that the strength of the non-linearity mainly depends on the frequency at which the non-linear effect takes place, $\omega_\alpha=\omega_{\alpha_0}$, and not on the frequencies contributing to the non-linear dynamics. This allows us to set the frequencies in the non-linear term to $\omega_{\alpha_0}$ (frequency of the driven mode) as well as the group velocity terms to $v_{\alpha_0}$. When the different frequencies are not independent and fulfill the energy conservation $\omega_\alpha+\omega_\eta=\omega_\mu+\omega_{\alpha+\eta-\mu}$, which is our case of interest, such approximation is expected to describe the non-linear phenomena accurately \cite{colman}. Equation~\ref{cme5} is thus turned into the following simpler expression
\begin{equation}\label{cme6}
 \dot{{\cal B}}_{\alpha}(t) + \left[\frac{\gamma_\alpha}{2}-i\sigma_\alpha\right]{\cal B}_{\alpha}(t)-\frac{il\omega_{\alpha_0}n_{g,\alpha_0}\tilde{n}_2}{\epsilon V_c}\sum_{\mu\eta}{\cal B}_{\mu}(t){\cal B}_{\eta}^\ast(t){\cal B}_{\alpha+\eta-\mu}(t)-\frac{\gamma_\alpha}{2}{\cal F}_\alpha\delta_{\alpha,\alpha_0}=0,
\end{equation}
where $n_{g,\alpha_0}=c/v_{\alpha_0}$ is the group index at $\omega_{\alpha_0}$. If we define the complex gain as
\begin{equation}\label{complexG}
 G_{\alpha_0}=g_{\alpha_0}+ig_{\alpha_0}^{\rm TPA}
\end{equation}
with
\begin{equation}\label{reg}
 g_{\alpha_0}=\frac{l\omega_{\alpha_0}n_{g,\alpha_0}n_2}{\epsilon V_c},
\end{equation}
being the real gain comming from the non-linear Kerr effect, and
\begin{equation}\label{img}
 g_{\alpha_0}^{\rm TPA}=\frac{lcn_{g,\alpha_0}\beta_{\rm TPA}}{2\epsilon V_c},
\end{equation}
being the \textit{imaginary gain}, representing the non-linear losses comming from TPA, we finally get the non-linear system of equations for the mode slowly-varying normal mode amplitudes in a CCW, with identical single mode cavities and periodic boundary conditions
\begin{equation}\label{cme7}
  \dot{{\cal B}}_{\alpha}(t) + \left[\frac{\gamma_\alpha}{2}-i\sigma_\alpha\right]{\cal B}_{\alpha}(t)-iG_{\alpha_0}\sum_{\mu\eta}{\cal B}_{\mu}(t){\cal B}_{\eta}^\ast(t){\cal B}_{\alpha+\eta-\mu}(t)-\frac{\gamma_\alpha}{2}{\cal F}_\alpha\delta_{\alpha,\alpha_0}=0.
\end{equation}
Equation~\ref{cme7} is equivalent to the one previously derived by Chembo and Yu in non-linear ring resonators \cite{chembo1}.

\section{Stationary solution and stability analysis}
In order to simplify the analysis of the stationary solutions of Eq.~\ref{cme7}  and their stability we introduce shifted momentum indices $\{\alpha',\mu',\eta'\}\rightarrow\{\alpha,\mu,\eta\}-\alpha_0$, where the driven mode has zero index and the side modes are symmetrically distributed around the driven one. In terms of this new indexes Eq.~\ref{cme7} is written as
\begin{equation}\label{cme7s}
  \dot{{\cal B}}_{\alpha'}(t) + \left[\frac{\gamma_{\alpha'}}{2}-i\sigma_{\alpha'}\right]{\cal B}_{\alpha'}(t)-iG_0\sum_{\mu'\eta'}{\cal B}_{\mu'}(t){\cal B}_{\eta'}^\ast(t){\cal B}_{\alpha'+\eta'-\mu'}(t)-\frac{\gamma_{\alpha'}}{2}{\cal F}_{\alpha'}\delta_{\alpha',0}=0.
\end{equation}

\subsection{Stationary solution}
The stationary solution of Eq.~\ref{cme7s} is found by setting the derivative term to zero
\begin{equation}\label{cme7sss}
  \left[\frac{\gamma_{\alpha'}}{2}-i\sigma_{\alpha'}\right]{\cal B}_{\alpha'}^s-iG_0\sum_{\mu'\eta'}{\cal B}_{\mu'}^s{\cal B}_{\eta'}^{s\ast}{\cal B}_{\alpha'+\eta'-\mu'}^s-\frac{\gamma_{\alpha'}}{2}{\cal F}_{\alpha'}\delta_{\alpha',0}=0.
\end{equation}

\subsection{System below threshold}
When the system is below threshold only the $\alpha'=0$ mode is excited, then, Eq.~\ref{cme7sss} becomes
\begin{equation}\label{centralss}
 \frac{\gamma_0}{2}{\cal B}_0^s-i\sigma_0{\cal B}_0^s-iG_0|{\cal B}_0^s|^2{\cal B}_0^s-\frac{\gamma_0}{2}{\cal F}_0=0,
\end{equation}
and we get the very well known cubic relation between $|{\cal F}_0|^2$ and $|{\cal B}_0^s|^2$ which leads to the hysteresis phenomenon and bistability of the stationary solution
\begin{equation}\label{hysteq0}
 |{\cal F}_0|^2=\left(1+\frac{4\sigma_0^2}{\gamma_0^2}\right)|{\cal B}_0^s|^2+\frac{4}{\gamma_0}\left(\frac{2g_0\sigma_0}{\gamma_0}+g_0^{\rm TPA}\right)|{\cal B}_0^s|^4+\frac{4|G_0|^2}{\gamma_0^2}|{\cal B}_0^s|^6.
\end{equation}
The bistability boundaries $B_\pm$ correspond to the local extrema of Eq.~\ref{hysteq0}, which are the solutions of the quadratic equation $\partial |{\cal F}_0|^2/\partial |{\cal B}_0^s|^2=0$ on $|{\cal B}_0^s|^2$. Thus, we readily obtain
\begin{equation}\label{Bbist}
 B_\pm=\frac{-\left(2\sigma_{0}g_0+\gamma_0g_0^{\rm TPA}\right) \pm \frac{1}{2}\sqrt{4\left(2\sigma_{0}g_0+\gamma_0g_0^{\rm TPA}\right)^2-3\left|G_0\right|^2\left(4\sigma_{0}^2+\gamma_0^2\right)}}{3\left|G_0\right|^2}.
\end{equation}
Hysteresis exists if the boundaries $B_\pm$ are real and positive, implying the following condition on the laser detuning
\begin{equation}\label{conshyst}
\sigma_{0}<\sigma_{\rm hyst}\rho(\kappa),
\end{equation}
where
\begin{equation}\label{shyst}
\sigma_{\rm hyst}=-\frac{\gamma_0\sqrt{3}}{2},
 \end{equation}
and $\rho(\kappa)$ is a function which depends on the material properties only
\begin{equation}\label{rhocrit}
 \rho(\kappa)=\frac{(4\sqrt{3}+3\kappa)\kappa+3}{3(1-3\kappa^2)},
\end{equation}
with the dimensionless parameter $\kappa$ defined as
\begin{equation}\label{kap}
 \kappa=\frac{g_{0}^{\rm TPA}}{g_{0}}=\frac{c\beta_{\rm TPA}}{2n_2\omega_{0}}.
\end{equation}
Notice that by setting $g_0^{\rm TPA}=0$ in Eqs.~\ref{hysteq0}, \ref{Bbist} and \ref{conshyst}, we recover exactly the same expressions for the ring resonator \cite{chembo1}.

\subsection{System at threshold}
The onset of the oscillations of a pair of side modes ${\cal B}_{\pm\alpha'}$ can be determined by studying the linear stability around the trivial equilibrium, i.e., ${\cal B}_{\pm\alpha'}^s=0$. This technique consists in adding a small time-dependent perturbation $\delta{\cal B}_{\pm\alpha'}(t)$ to the stationary equilibrium, and the set of parameters leading to the divergence of $\delta{\cal B}_{\pm\alpha'}(t)$ defines the threshold for side mode oscillations. Hence, the mode amplitude ${\cal B}_{\alpha'}(t)$ is written as
\begin{equation}\label{Btrivial}
{\cal B}_{\alpha'}(t)={\cal B}_{\alpha'}^s\delta_{\alpha',0}+\delta{\cal B}_{\alpha'}(t),
\end{equation}
and replaced in Eq.~\ref{cme7sss}, which gives 
\begin{align}
\frac{d\delta{\cal B}_{\alpha'}(t)}{dt} = & -\left[\frac{\gamma_{\alpha'}}{2}-i\sigma_{\alpha'}\right]\delta{\cal B}_{\alpha'}(t) \nonumber \\
&+iG_0\sum_{\mu'\eta'}\left[{\cal B}_{\mu'}^s\delta_{\mu',0}+\delta{\cal B}_{\mu'}(t)\right]\left[{\cal B}_{\eta'}^s\delta_{\eta',0}+\delta{\cal B}_{\eta'}(t)\right]^\ast\left[{\cal B}_{\alpha'+\eta'-\mu'}^s\delta_{\alpha'+\eta'-\mu',0}+\delta{\cal B}_{\alpha'+\eta'-\mu'}(t)\right] \nonumber \\
&-\left[\frac{\gamma_{\alpha'}}{2}-i\sigma_{\alpha'}\right]{\cal B}_{\alpha'}^s\delta_{\alpha',0} + \frac{\gamma_{\alpha'}}{2}{\cal F}_{\alpha'}\delta_{\alpha',0},
\end{align}
where we have written explicitly the time derivative operator for the sake of clarity. By keeping only terms linear in $\delta{\cal B}$, and using the steady state solution of the central mode in Eq.~\ref{centralss}, we get
\begin{equation}\label{stabeq1}
\frac{d\delta{\cal B}_{\alpha'}(t)}{dt} = -\left[\frac{\gamma_{\alpha'}}{2}-i\sigma_{\alpha'}\right]\delta{\cal B}_{\alpha'}(t)+iG_0\left[2|{\cal B}_0^s|^2\delta{\cal B}_{\alpha'}(t)+ {{\cal B}_0^{s}}^2\delta{\cal B}_{-\alpha'}^\ast(t)\right]    
\end{equation}
For studying the dynamical behavior of the time-dependent perturbations, Eq.~\ref{stabeq1} can be rewritten in a more convenient form by making the transformations 
\begin{align}\label{stabTransf}
 {\cal C}_0^s = & {\cal B}_0^s \nonumber \\
 \delta{\cal C}_{\alpha'}(t) = & \delta{\cal B}_{\alpha'}(t)e^{i\left(\sigma_0-\sigma_{\alpha'}+\frac{1}{2}\overline{\omega}_{\alpha'}\right)t},  
\end{align}
with
\begin{equation}\label{disp}
 \overline{\omega}_{\alpha'}=2\omega_0-\omega_{\alpha'}-\omega_{-\alpha'}.
\end{equation}
Moreover, in order to simplify the stability analysis, we replace the loss rates of the side modes $\gamma_{\pm\alpha'}$ by $\gamma_0$, which is a good approximation when $\gamma$ does not have strong fluctuations within the interval $[-\alpha',+\alpha']$. Equation~\ref{stabeq1} is finally turned into 
\begin{equation}\label{stabeq2}
\frac{d\delta{\cal C}_{\alpha'}(t)}{dt} = \left[i\left(\sigma_0+\frac{1}{2}\overline{\omega}_{\alpha'}\right)-\frac{\gamma_0}{2}\right]\delta{\cal C}_{\alpha'}(t)+iG_0\left[2|{\cal C}_0^s|^2\delta{\cal C}_{\alpha'}(t)+ {{\cal C}_0^{s}}^2\delta{\cal C}_{-\alpha'}^\ast(t)\right].    
\end{equation}
Notice that the coefficients of Eq.~\ref{stabeq2} are invariant to the change $\alpha' \rightarrow -\alpha'$ because $\overline{\omega}_{\alpha'}=\overline{\omega}_{-\alpha'}$, therefore, the dynamics of the side mode perturbations are described by the following two linearized equations
\begin{align}\label{linstaeqs}
 \frac{d\delta{\cal C}_{\alpha'}(t)}{dt} = & M_1\delta{\cal C}_{\alpha'}(t)+M_2\delta{\cal C}_{-\alpha'}^\ast(t) \nonumber \\
 \frac{d\delta{\cal C}_{-\alpha'}^\ast(t)}{dt} = & M_2^\ast\delta{\cal C}_{\alpha'}(t)+M_1^\ast\delta{\cal C}_{-\alpha'}^\ast(t),
\end{align}
where
\begin{align}\label{Mcoeff}
 M_1 = & -\frac{\gamma_0}{2}+i\left(\sigma_0+\frac{1}{2}\overline{\omega}_{\alpha'}\right)+2iG_0|{\cal C}_0^s|^2 \nonumber \\
 M_2 = & iG_0{{\cal C}_0^{s}}^2.
\end{align}
The stability of the trivial equilibrium is then assessed by analyzing if the perturbation decays to zero, which means that such equilibrium is stable and there are no side mode oscillations, or if the perturbation diverges, meaning that the trivial equilibrium is unstable thus leading to the onset of side mode oscillations. This is carried out by writing 
\begin{align}\label{Clam}
 \delta{\cal C}_{\alpha'}(t) = & e^{\lambda t}\delta{\cal C}'_{\alpha'} \nonumber \\
 \delta{\cal C}_{-\alpha'}^\ast(t) = & e^{\lambda t}\delta{\cal C}_{-\alpha'}'^{\ast},
\end{align}
which together with Eq.~\ref{linstaeqs} sets the following $2\times2$ eigenvalue problem
\begin{equation}\label{eigenlam}
 \begin{bmatrix}
  M_1 & M_2 \\
  M_2^\ast & M_1^\ast
 \end{bmatrix}
 \begin{bmatrix}
  \delta{\cal C}'_{\alpha'} \\
  \delta{\cal C}_{-\alpha'}'^{\ast}
 \end{bmatrix} = \lambda
 \begin{bmatrix}
  \delta{\cal C}'_{\alpha'} \\
  \delta{\cal C}_{-\alpha'}'^{\ast}
 \end{bmatrix}.
\end{equation}
The solutions for $\lambda$ in Eq.~\ref{eigenlam} are readily found
\begin{align}\label{lamsol}
\lambda_\pm = {\rm Re}\left\{M_1\right\}\pm\sqrt{\left|M_2\right|^2-{\rm Im}\left\{M_1\right\}^2} 
= \frac{\gamma_0}{2}-2{g_{0}^{\rm TPA}}^2\left|{\cal C}_0^s\right|^2\pm\sqrt{\left({g_{0}^{\rm TPA}}^2-3g_0^2\right)\left|{\cal C}_0^s\right|^4-4g_0\left|{\cal C}_0^s\right|^2\Delta_{\alpha'}-\Delta_{\alpha'}^2},
\end{align}
where we have defined 
\begin{equation}\label{sidedet}
\Delta_{\alpha'}=\sigma_0+\frac{1}{2}\overline{\omega}_{\alpha'}.
\end{equation}
The divergence of Eq.~\ref{Clam}, i.e., the onset of the side mode oscillations, is clearly obtained when ${\rm Re}\{\lambda_+\}>0$, which in terms of the system parameters reads 
\begin{equation}\label{Sstab}
{\cal S}_{\alpha'}=12|G_0|^2|{\cal C}_0^s|^4+8\left(2\Delta_{\alpha'}g_0+\gamma_0g_0^{\rm TPA}\right)|{\cal C}_0^s|^2+4\Delta_{\alpha'}^2+\gamma_0^2<0.
\end{equation}
The zeros of Eq.~\ref{Sstab} determine the boundaries in which the trivial equilibrium becomes unstable. Therefore, by defining $\tilde{B}=|{\cal C}_0^s|^2$, these boundaries are the solutions of the quadratic equation ${\cal S}_{\alpha'}(\tilde{B})=0$, i.e., 
\begin{equation}\label{Bbounds}
 \tilde{B}_\pm=\frac{-\left(2\Delta_{\alpha'}g_0+\gamma_0g_0^{\rm TPA}\right) \pm \frac{1}{2}\sqrt{4\left(2\Delta_{\alpha'}g_0+\gamma_0g_0^{\rm TPA}\right)^2-3\left|G_0\right|^2\left(4\Delta_{\alpha'}^2+\gamma_0^2\right)}}{3\left|G_0\right|^2}.
\end{equation}
Notice that Eqs.~\ref{Sstab} and \ref{Bbounds} reduce exactly to the same corresponding expressions reported in Ref.~\cite{chembo1} for $g_0^{\rm TPA}=0$. The minimum mode power leading to side mode oscillation is determined by the equation 
\begin{equation}\label{thcond}
\frac{\partial \tilde{B}_-}{\partial \Delta_{\alpha'}}=0.
\end{equation}
In terms of the fundamental CCW parameters and the actual momentum index (no shifted), this modal threshold power for comb generation is found to be
\begin{equation}\label{Th}
\left|{\cal A}_{\alpha_0}\right|_{\rm th}^2=\frac{\gamma_{\alpha_0}}{2g_{\alpha_0}}f(\kappa)=\frac{\epsilon V_c}{2ln_{g,\alpha_0}n_2Q_{\alpha_0}}f(\kappa),
\end{equation}
where $Q_{\alpha_0}=\omega_{\alpha_0}/\gamma_{\alpha_0}$ is the quality factor of the driven CCW normal mode $\alpha_0$, and $f(\kappa)$ is a function of $\kappa$ only
\begin{equation}\label{fdef}
 f(\kappa)=\frac{\sqrt{1+\kappa^2}+2\kappa}{1-3\kappa^2},
\end{equation}
From Eqs.~\ref{Th} and \ref{fdef} we clearly see that, as far as the threshold is concerned, TPA increases the minimum power required to start comb generation. The corresponding driven amplitude threshold $|{\cal F}_{\alpha_0}|_{\rm th}^2(\sigma_{\alpha_0})$, which depends on the frequency detuning between the external laser and driven mode, is obtained by replacing Eq.~\ref{Th} on Eq.~\ref{hysteq0} and recalling that $|{\cal A}|=|{\cal B}|$, thus
\begin{equation}\label{Fth}
|{\cal F}_{\alpha_0}|_{\rm th}^2(\sigma_{\alpha_0})=\frac{2\sigma_{\alpha_0}^2}{g_{\alpha_0}\gamma_{\alpha_0}}f(\kappa)+\frac{2\sigma_{\alpha_0}}{g_{\alpha_0}}f^2(\kappa)+\frac{\gamma_{\alpha_0}}{2g_{\alpha_0}}f(\kappa)+\frac{g_{\alpha_0}^{\rm TPA}\gamma_{\alpha_0}}{g_{\alpha_0}^2}f^2(\kappa)+\frac{|G_{\alpha_0}|^2\gamma_{\alpha_0}}{2g_{\alpha_0}^3}f^3(\kappa).
\end{equation}
The optimal detuning that leads to the absolute minimal driven amplitude, required by comb generation, is easily found by solving the equation $\partial|{\cal F}_{\alpha_0}|_{\rm th}^2/\partial\sigma_{\alpha_0}=0$, and it is given by
\begin{equation}\label{optdet}
\sigma_{\alpha_0}^{\rm th}=-\frac{\gamma_{\alpha_0}}{2}f(\kappa),
\end{equation}
Equation~\ref{optdet} shows that in presence of TPA, the optimal laser detuning to start frequency combs is red-shifted.\\

\noindent A closer look to Eq.~\ref{Th} evidences that $\left|{\cal A}_{\alpha_0}\right|_{\rm th}^2$ depends actually on the effective area of the cavity mode
\begin{equation}\label{ThAeff}
 \left|{\cal A}_{\alpha_0}\right|_{\rm th}^2=\frac{\epsilon A_{\rm eff}}{2n_{g,\alpha_0}n_2Q_{\alpha_0}}f(\kappa),
\end{equation}
where the mode area $A_{\rm eff}=V_c/l$ is widely employed in waveguide physics. Moreover, if we consider the limit of no TPA, i.e., $f(\kappa=0)=1$, and the limit of no internal nanostrcuture (homogeneous slab for photonic crystals) where the group index is very close to the actual refractive index of the dielectric $n$, i.e., $\epsilon=n^2\backsimeq n_{g,\alpha_0}^2$ , Eq.~\ref{ThAeff} becomes
\begin{equation}\label{ThAeffhom}
 \left|{\cal A}_{\alpha_0}\right|_{\rm th}^2=\frac{n A_{\rm eff}}{2n_2Q_{\alpha_0}},
\end{equation}
which is exactly the same threshold expression (in Watts) for the internal mode power in a non-linear ring resonator \cite{chembo2}.

\subsection{Role of dispersion in frequency comb generation}
In order to the boundaries of Eq.~\ref{Bbounds} be real and positive, the discriminant of the square root has to be positive and the term $\left(2\Delta_{\alpha'}g_0+\gamma_0g_0^{\rm TPA}\right)$ has to be negative. Both conditions are satisfied if
\begin{equation}\label{conscrit}
\Delta_{\alpha'}<\sigma_{\rm cr}\rho(\kappa),
\end{equation}
where the critical detuning $\sigma_{\rm cr}$ is defined as
\begin{equation}\label{scrit}
\sigma_{\rm cr}=-\frac{\gamma_0\sqrt{3}}{2},
 \end{equation}
If the dispersion relation $\omega_{\alpha'}$ is expanded in Taylor series around $\omega_0$ up to third order, we have
\begin{equation}\label{wexp3ord}
 \omega_{\alpha'}=\omega_0+\zeta_1{\alpha}'+\frac{\zeta_2}{2}{\alpha'}^2+\frac{\zeta_3}{3!}{\alpha'}^3,
\end{equation}
with $\zeta_1$, $\zeta_2$ and $\zeta_3$ representing the group velocity, group velocity dispersion and third order dispersion, respectively. Replacing Eq.~\ref{wexp3ord} in Eq.~\ref{disp} gives
\begin{equation}\label{disp2}
 \overline{\omega}_{\alpha'}=-\zeta_2{\alpha'}^2,
\end{equation}
turning the condition of Eq.~\ref{conscrit} into
\begin{equation}\label{conscrit2}
-\zeta_2{\alpha'}^2<2\left[\sigma_{\rm cr}\rho(\kappa)-\sigma_0\right].
\end{equation}
For normal dispersion $\zeta_2<0$ and Eq.~\ref{conscrit2} defines the following condition for $\alpha'$ 
\begin{equation}\label{normaldis}
 \alpha'<\sqrt{\frac{2}{|\zeta_2|}\left[\sigma_{\rm cr}\rho(\kappa)-\sigma_0\right]}=\alpha'_{\rm max},
\end{equation}
while for anomalous dispersion $\zeta_2>0$ the condition determined by Eq.~\ref{conscrit2} reads
\begin{equation}\label{anomalousdis}
 \alpha'>\sqrt{\frac{2}{|\zeta_2|}\left[\sigma_0-\sigma_{\rm cr}\rho(\kappa)\right]}=\alpha'_{\rm min}.
\end{equation}
Therefore, from Eqs.~\ref{normaldis} and \ref{anomalousdis}, we clearly see that if the system is pumped  where dispersion is normal, there is an upper bound on the side mode momentum, thus constraining the spanning of the frequency comb. Nevertheless, if the dispersion is anomalous, there is no upper bound on the momentum for side mode oscillations, and the comb generation regime is easily achieved.

\section{Lugiato-Lefever equation}
The corresponding spatio-temporal equation for the non-linear formalism presented in Sec.~\ref{cmesec} can be obtained by following the derivation of Ref.~\cite{chembo3}. We start from the system of coupled mode equations in Eq.~\ref{cme7}, which can be rewritten as
\begin{equation}\label{cme72}
  \dot{{\cal B}}_{\alpha}(t) + \left[\frac{\gamma_\alpha}{2}-i\sigma_\alpha\right]{\cal B}_{\alpha}(t)-ig_{\alpha_0}(1+i\kappa)\sum_{\mu\eta\xi}{\cal B}_{\mu}(t){\cal B}_{\eta}^\ast(t){\cal B}_{\xi}(t)\delta_{\xi,\alpha+\eta-\mu}-\frac{\gamma_\alpha}{2}{\cal F}_\alpha\delta_{\alpha,\alpha_0}=0,
\end{equation}
where we have explicitly considered the delta factor in the non-linear term and used Eqs.~\ref{complexG}, \ref{reg}, \ref{img} and \ref{kap} to write $G_{\alpha_0}=g_{\alpha_0}(1+i\kappa)$. We define the following spatio-temporal slowly-varying envelope along the waveguide direction
\begin{equation}\label{Lenvelope}
\psi(y,t)=\sum_\alpha{\cal B}_{\alpha}(t)e^{-i(\alpha-\alpha_0)y},    
\end{equation}
from which we have
\begin{align}
    \frac{\partial\psi(y,t)}{\partial t} = & \sum_\alpha\dot{{\cal B}}_{\alpha}(t)e^{-i(\alpha-\alpha_0)y},\label{tderiv}\\
    i^m\frac{\partial^m\psi(y,t)}{\partial y^m} = & \sum_\alpha(\alpha-\alpha_0)^m{\cal B}_{\alpha}(t)e^{-i(\alpha-\alpha_0)y}. \label{sderiv}
\end{align}
Equation~\ref{cme72} is now replaced in Eq.~\ref{tderiv} leading to
\begin{align}\label{tderiv2}
     \frac{\partial\psi(y,t)}{\partial t} = \sum_\alpha\left[i\sigma_\alpha-\frac{\gamma_\alpha}{2}\right]{\cal B}_{\alpha}(t)e^{-i(\alpha-\alpha_0)y} 
     + ig_{\alpha_0}(1+i\kappa)\sum_{\mu\eta\xi}{\cal B}_{\mu}(t)e^{-i(\mu-\alpha_0)y}{\cal B}_{\eta}^\ast(t)e^{i(\eta-\alpha_0)y}{\cal B}_{\xi}(t)e^{-i(\xi-\alpha_0)y}+\frac{\gamma_{\alpha_0}}{2}{\cal F}_{\alpha_0},
\end{align}
and because 
\begin{equation}\label{psi3}
|\psi(y,t)|^2\psi(y,t)=\sum_{\mu\eta\xi}{\cal B}_{\mu}(t)e^{-i(\mu-\alpha_0)y}{\cal B}_{\eta}^\ast(t)e^{i(\eta-\alpha_0)y}{\cal B}_{\xi}(t)e^{-i(\xi-\alpha_0)y},
\end{equation}
we readily get from Eq.~\ref{tderiv2}
\begin{equation}\label{tderiv3}
\frac{\partial \psi}{\partial t}=\sum_\alpha\left[i\sigma_\alpha-\frac{\gamma_\alpha}{2}\right]{\cal B}_{\alpha}(t)e^{-i(\alpha-\alpha_0)y} + ig_{\alpha_0}(1+i\kappa) |\psi|^2\psi +\frac{\gamma_{\alpha_0}}{2}{\cal F}_{\alpha_0},
\end{equation}
where we have simplified the notation $\psi(y,t)\rightarrow\psi$. By writing the mode detuning as
\begin{equation}\label{moddet}
\sigma_\alpha=\Omega_0-\omega_\alpha=\Omega_0-\omega_{\alpha_0}-(\omega_\alpha-\omega_{\alpha_0})=\sigma_{\alpha_0}-(\omega_\alpha-\omega_{\alpha_0}),
\end{equation}
and expanding the last term in Taylor series around $\omega_{\alpha_0}$
\begin{equation}\label{taylorw}
\omega_\alpha-\omega_{\alpha_0}=\sum_m\frac{\zeta_m}{m!}(\alpha-\alpha_0)^m,    
\end{equation}
Eq.~\ref{tderiv3} is turned into
\begin{align}\label{tderiv4}
\frac{\partial \psi}{\partial t}  = & ig_{\alpha_0}(1+i\kappa) |\psi|^2\psi + i\sigma_{\alpha_0}\sum_\alpha{\cal B}_{\alpha}(t)e^{-i(\alpha-\alpha_0)y}-\sum_\alpha\frac{\gamma_\alpha}{2}{\cal B}_{\alpha}(t)e^{-i(\alpha-\alpha_0)y} \nonumber \\
&-i\sum_m\frac{\zeta_m}{m!}\sum_\alpha (\alpha-\alpha_0)^m{\cal B}_{\alpha}(t)e^{-i(\alpha-\alpha_0)y}+\frac{\gamma_{\alpha_0}}{2}{\cal F}_{\alpha_0}.
\end{align}
If we use Eqs.~\ref{Lenvelope} and \ref{sderiv}, and assume an overall losses given by the loss rate at the pump frequency, i.e., $\gamma_\alpha=\gamma_{\alpha_0}$ \footnote{We are ultimately assuming that the loss rates do not depend on the frequency or, equivalently, on the momentum. Such losses may be given by $\gamma_{\alpha_0}$ or by any reliable constant value within the Brillouin zone of the CCW.}, Eq.~\ref{tderiv4} takes the following form
\begin{equation}\label{LLEq1}
\frac{\partial \psi}{\partial t} =  \left(i\sigma_{\alpha_)}-\frac{\gamma_{\alpha_0}}{2}\right)\psi+ ig_{\alpha_0}(1+i\kappa) |\psi|^2\psi +\zeta_1\frac{\partial \psi}{\partial y}+i\frac{\zeta_2}{2}\frac{\partial^2\psi}{\partial y^2}+\sum_{m=3}i^{m-1}\frac{\zeta_m}{m!}\frac{\partial^m\psi}{\partial y^m}+\frac{\gamma_{\alpha_0}}{2}{\cal F}_{\alpha_0},
\end{equation}
where the group velocity $\zeta_1$ and the group velocity dispersion $\zeta_2$ have been written explicitly. In order to get rid of the first spatial-derivative term we change to a moving reference frame with velocity $\zeta_1$, i.e., 
\begin{align}\label{tranformyt}
y'=  &  y+\zeta_1 t \nonumber \\
t'=  &  t 
\end{align}
which gives 
\begin{align}\label{tranformder}
\frac{\partial}{\partial t} = & \frac{\partial}{\partial t'} + \zeta_1\frac{\partial}{\partial y'}, \nonumber\\
\frac{\partial}{\partial y} = & \frac{\partial}{\partial y'}.
\end{align}
Replacing Eq.~\ref{tranformder} in Eq.~\ref{LLEq1}, and introducing the normalized detuning $\varsigma=-2\sigma_{\alpha_0}/\gamma_{\alpha_0}$, we arrive to
\begin{equation}\label{LLEq2}
\frac{\partial \psi}{\partial t'} =-(1+i\varsigma)\frac{\gamma_{\alpha_0}}{2}\psi+ig_{\alpha_0}(1+i\kappa) |\psi|^2\psi+i\frac{\zeta_2}{2}\frac{\partial^2\psi}{\partial y'^2}+\sum_{m=3}i^{m-1}\frac{\zeta_m}{m!}\frac{\partial^m\psi}{\partial y'^m}+\frac{\gamma_{\alpha_0}}{2}{\cal F}_{\alpha_0}.  
\end{equation}
By defining the quantities $\Psi = \sqrt{2g_{\alpha_0}/\gamma_{\alpha_0}}\psi$, $F =  \sqrt{2g_{\alpha_0}/\gamma_{\alpha_0}}{\cal F}_{\alpha_0}$, $\varrho_m = -2\zeta_m/\gamma_{\alpha_0}$ and $\tau = (\gamma_{\alpha_0}/2)t'$ we finally get the normalized Lugiato-Lefever equation with non-linear losses (TPA) and high-order dispersion
\begin{equation}
\frac{\partial\Psi}{\partial\tau}=-(1+i\varsigma)\Psi+i(1+i\kappa)|\Psi|^2\Psi-i\frac{\varrho_2}{2}\frac{\partial^2\Psi}{\partial y'^2} + \sum_{m=3}i^{m+1}\frac{\varrho_m}{m!}\frac{\partial^m\Psi}{\partial y'^m}+F 
\end{equation}

\newpage

\section{Soliton dynamics}

\begin{figure}[h!]
\centering\includegraphics[width=0.9\textwidth]{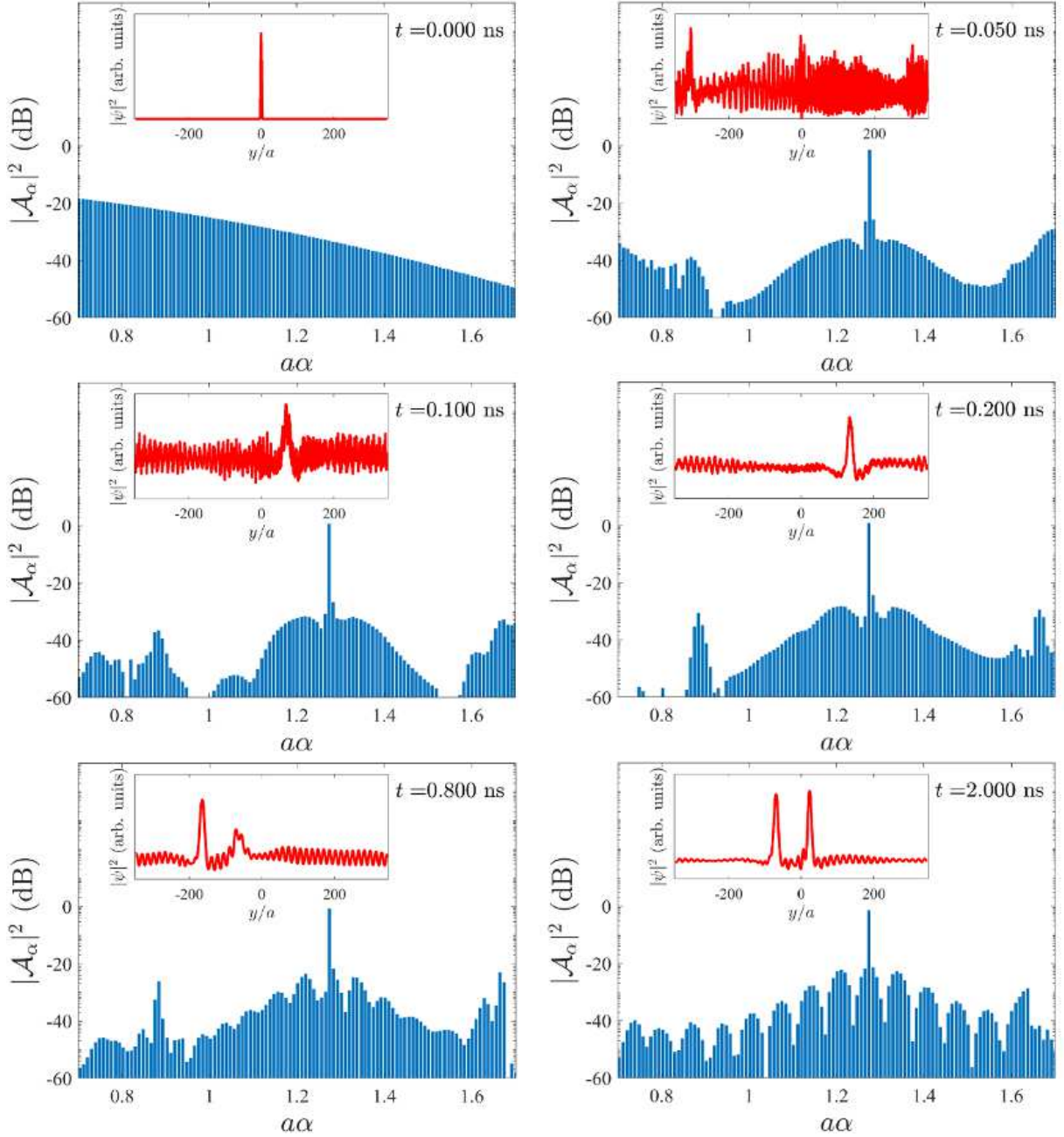}
\caption{Transient dynamics of the spectral components for the molecule soliton of two pulses. The corresponding envelope function is shown in the inset. $|{\cal A}_\alpha|^2$ is given in threshold units and the initial condition ($t=0$) is the sharp Gaussian pulse $\psi(y,t=0)=\exp[-0.5(y/\sqrt{3}a)^2]$.}\label{figsolIII}
\end{figure}

\begin{figure}[t!]
\centering\includegraphics[width=0.9\textwidth]{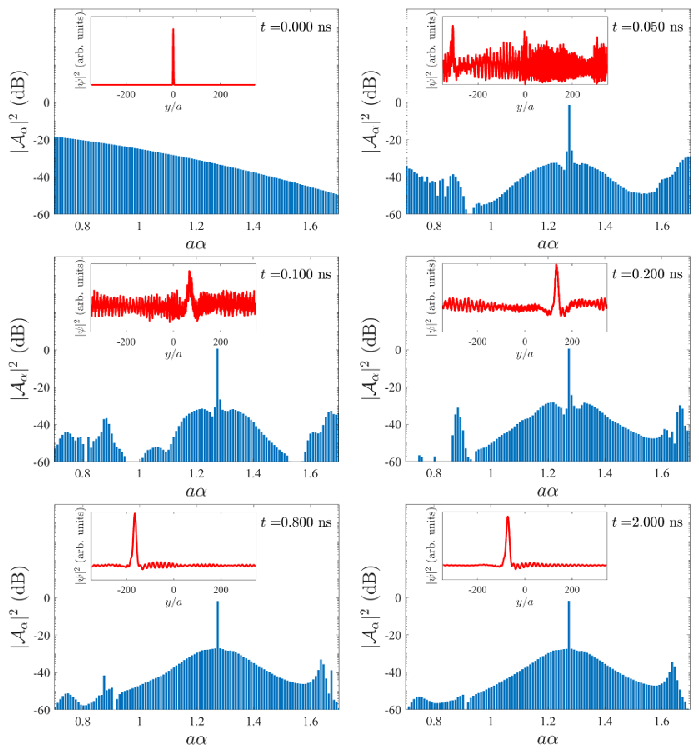}
\caption{Transient dynamics of the spectral components for a single soliton pulse. The corresponding envelope function is shown in the inset. $|{\cal A}_\alpha|^2$ is given in threshold units and the initial condition ($t=0$) is the sharp Gaussian pulse $\psi(y,t=0)=\exp[-0.5(y/\sqrt{3}a)^2]$.}\label{figsolIV}
\end{figure}

We report in Figs.~\ref{figsolIII}~and~\ref{figsolIV} the spectral dynamics of soliton formation for the two relevant cases III and IV, respectively, presented in Figs.~2~and~3 in the manuscript. Panels correspond to different times (in ns units) during the transient dynamics and the associated envelope function along the waveguide direction is shown in the inset.

\newpage

\twocolumngrid


%

\end{document}